\documentclass[aps,prl,amsmath,amssymb,twocolumn,floatfix]{revtex4-1}
\usepackage{graphicx}
\usepackage{dcolumn}
\usepackage{bm}

\begin{document}

\title{Long Range Magnetic Order in the Anisotropic Triangular Lattice System CeCd$_3$As$_3$}

\author{S. R.~Dunsiger,$^{1,2}$ J.~Lee,$^1$ J. E.~Sonier,$^{1,3}$ E. D. Mun$^{1}$}

\affiliation{$^1$Department of Physics, Simon Fraser University, Burnaby, British Columbia, Canada V5A 1S6 \\
$^2$Centre for Molecular and Materials Science, TRIUMF, Vancouver, British Columbia, Canada, V6T 2A3 \\
$^3$Canadian Institute for Advanced Research, Toronto, Ontario, Canada M5G 1Z8}

\date{\today}
\begin{abstract}
We report the physical properties of $R$Cd$_{3}$As$_{3}$ ($R$ = La and Ce) compounds, crystallized into a hexagonal ScAl$_{3}$C$_{3}$-type structure ($P$6$_{3}$/mmc) such that the $R$ sublattice forms a spin-orbit coupled triangular lattice. Magnetic susceptibility measurements indicate the 4$f$ electrons of Ce$^{3+}$ ions are well localized and reveal a large magnetic anisotropy. The electrical resistivity and specific heat measurement for $R$Cd$_{3}$As$_{3}$ exhibit an anomaly at high temperatures ($T_{0}$ $\sim$ 63 K for $R$ = La and $T_{0}$ $\sim$ 136 K for $R$ = Ce), most likely due to a structural transition. Specific heat measurements for CeCd$_{3}$As$_{3}$ clearly indicate a long range magnetic order below $T_{N}$ = 0.42 K. Although the magnetic contribution to the specific heat $C_{m}/T$ increases significantly below $\sim$ 10 K, the electrical resistivity for CeCd$_{3}$As$_{3}$ follows typical, metallic behavior inconsistent with Kondo lattice systems. In CeCd$_3$As$_3$ only $\sim$ 40 $\%$ of the $R \ln(2)$ magnetic entropy is recovered by $T_N$ and the $R$ln(2) entropy is fully achieved at about the Curie-Weiss temperature $|\theta_{p}|$. Unusually, based on our current investigations, the magnetic specific heat below $|\theta_{p}|$ is not attributed to a Kondo contribution, but rather associated with the magnetic ordering and frustration on  the triangular lattice.  Specific heat measurements in applied magnetic field show a negligible variation of $T_{N}$ for $H \parallel c$, whereas a suppression of $T_{N}$ is observed above 40 kOe for $H \parallel ab$. Such behavior is consistent with the application a magnetic field within the $ab$-plane breaking the triangular symmetry and partially relieving the magnetic frustration in this system.
\end{abstract}

\maketitle
\section{Introduction}

The phenomenon of geometric magnetic frustration is ubiquitous within condensed matter.  In such a scenario, it is not possible to simultaneously minimize all interactions in a system of interest, which offers a means of accessing novel ground states. Historically, spin 1/2 Ising \cite{wannier50} and Heisenberg \cite{anderson73} systems on a triangular lattice with nearest neighbor antiferromagnetic exchange interactions are the archetypal examples of geometric magnetic frustration. The triangular lattice systems were the first to be proposed as candidates for a quantum spin liquid ground state \cite{balents10}, where there is an unusually high degree entanglement of the spins and consequently the system is capable of supporting non-local excitations without ordering \cite{savary17}. The present theoretical consensus is that the ground state of a two dimensional triangular lattice has an ordered noncollinear 120$^{\circ}$ spin structure with a significantly reduced sublattice magnetization for both classical (large spin S) and quantum (S = 1/2) systems \cite{huse88,jolicoeur89,bernu94,singh92,white07}. More generally, the role of in-plane anisotropy and interplanar coupling must be included \cite{fazekas74,yamamoto12,schmidt14,li16}. Such anisotropy is particularly relevant to materials based on 4$f$ ions, where spin orbit effects become increasingly important \cite{krempa14} and the coupling between spins depends on bond orientation \cite{nussinov15}. 

Among rare-earth systems, there have been extensive investigations of Ce$^{3+}$ (4$f^{1}$) and Yb$^{3+}$ (4$f^{13}$) based intermetallic compounds, where rich phenomenology has been observed, including heavy Fermion behavior, unconventional superconductivity, and non-Fermi liquid behavior in the proximity of a quantum critical point \cite{sachdev01, vojta03, stewart06, lohneysen07, amato97}. The ground states of these systems are considered to be mainly governed by the competition between the Kondo and the Ruderman - Kittel - Kasuya - Yosida (RKKY) exchange interaction, both mediated by the conduction electrons. 
In addition to the energy scales associated with the separation of the 4$f$ -electron energy level from the Fermi level and the hybridization strength between the 4$f$ and the conduction electrons, the characteristic magnitude of the crystalline electric field (CEF) energy level splittings is also relevant\cite{bauer91}.  

Combining these two arenas, whereas earlier investigations concentrated on insulating systems, there are an increasing number of studies of frustrated magnetism in metallic systems \cite{batista16}, where the short-range interactions (nearest neighbor and next nearest neighbor) must be considered, in addition to the Kondo and RKKY interaction. Because the RKKY interaction is of a longer range than the superexchange interaction, the frustration effect in metallic compounds may be most evident in low carrier density systems. When quantum fluctuations become enhanced due to magnetic frustration, exotic phases such as complex spin density wave ordering or quantum spin liquids have been proposed \cite{coleman10, si10, coleman10a, kim13, mun13, tokiwa13, fritsch14}.  It is therefore clearly important to search for experimental realizations of such spin liquid behavior - rare-earth based triangular lattice compounds are  promising candidates.

Within such rare-earth triangular lattice systems, spin liquid behavior has been proposed in insulating YbMgGaO$_{4}$ \cite{li15a,li15b,paddison17}. The low carrier density system YbAl$_{3}$C$_{3}$ \cite{hara12} is thought to dimerize without ordering, opening up a spin gap in the magnetic excitation spectrum.  Recently, $RM_{3}X_{3}$ ($R$ = rare earth, $M$ = Al, Cd, and Zn, $X$ = C, P, and As) compounds, which crystallize in the hexagonal ScAl$_{3}$C$_{3}$-type structure (space group $P$6$_{3}$/mmc) have been synthesized and their physical properties reported \cite{nientiedt99,ochiai10,stoyko11}. In this crystal structure the magnetic rare-earth ions form two dimensional triangular lattice layers, well isolated by the interleaved $M$- and $X$-atoms. As such, they are good experimental realizations of anisotropic spin-orbit coupled triangular lattice models, where the role of conduction electrons may be systematically altered with lattice constant or pressure \cite{yamada10}. As anticipated, increasing rare-earth atomic number reduces the lattice constant, while increasing $X$ ion size acts to increase it \cite{nientiedt99}. The compounds become progressively more insulating with increasing unit cell volume: the $R$Al$_3$C$_3$ series is metallic in character \cite{ochiai10} (the $R$M$_3$As$_3$ series has a comparable cell volume), while CeZn$_3$P$_3$ and CeCd$_3$P$_3$ powder are reported to be semiconductors \cite{yamada10,higuch16}.  However, there are discrepancies - a recent investigation of single crystalline CeCd$_3$P$_3$ exhibits a conventional metallic response in its resistivity \cite{lee19}. 

We have successfully grown single crystals of CeCd$_3$As$_3$ and LaCd$_3$As$_3$. The isolated Ce$^{3+}$ ion is described in terms of a $J = 5/2$ ground state ($S = 1/2, L = 3$), further split into three Kramers doublets by the crystalline electric field (CEF). Hence it could in principle be described as an effective spin-1/2 ion. More detailed analysis \cite{banda18}, however, indicates significant mixing between $| \pm \frac{5}{2}>$ and  $| \mp \frac{1}{2}>$ states. The La-based compound is nonmagnetic, as the La$^{3+}$ ion has no unfilled electronic orbitals.  As such, a comparison of the two isostructural compounds allows us to disentangle structural and magnetic phenomena. In this article, we report the results of magnetic susceptibility, magnetization, electrical and Hall resistivity, and specific heat measurements of these compounds. The combination enables us to gain insight into the relative contributions from RKKY, Kondo and short-range interactions.

\section{Experimental}

\begin{figure}
\centering
\includegraphics[width=1\linewidth]{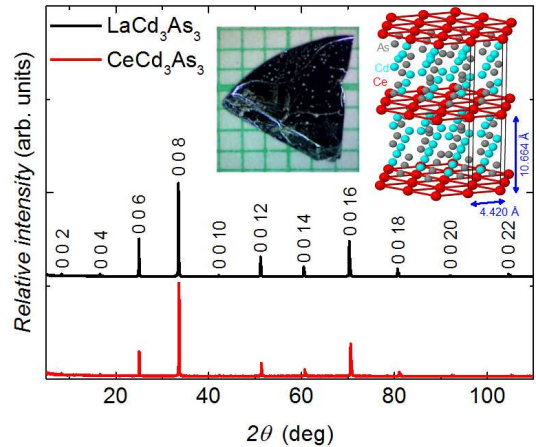}
\caption{Single crystal X-ray diffraction patterns for LaCd$_{3}$As$_{3}$ (top) and CeCd$_{3}$As$_{3}$ (bottom). The insets show a photograph of as grown LaCd$_{3}$As$_{3}$ over the mm scale and the crystal structure of $R$Cd$_{3}$As$_{3}$ ($R$ = La and Ce).}
\label{Fig1}
\end{figure}

Single crystals of $R$Cd$_3$As$_3$ ($R$ = La and Ce) were grown out of a ternary melt with excess As and Cd. The starting elements of high purity Ce, Cd, and As, taken in the ratio 1 : 10 : 10, were placed in an alumina crucible and sealed in a silica ampoule under a partial Ar pressure. The ampoule was heated slowly to 1050$^{o}$C in a box furnace and then cooled to 850$^{o}$C over 100 hours. After removing the excess amount of liquid by centrifuging, shiny platelet single crystals were obtained (Fig. \ref{Fig1}). The crystal structure of the samples was confirmed by powder X-ray diffraction (XRD) at room temperature. The peak positions in power XRD can be indexed with a hexagonal ScAl$_3$C$_3$-type structure ($P$6$_{3}$/mmc), consistent with previous reports \cite{stoyko11, liu16}. A large $c/a$ ratio suggests the most relevant interactions are those within the $ab$-plane. The grown single crystals form very thin plates, reflecting their layered structure.  The observation of only (0, 0, $l$) peaks, as shown in Fig. \ref{Fig1}, confirms the crystallographic $c$-axis is perpendicular to the plane of the thin plates.

The dc magnetization was measured in a Quantum Design (QD) Magnetic Property Measurement System (MPMS) as a function of temperature from 1.8 to 350~K, and magnetic field, up to 70 kOe. The specific heat, $C_{p}$, was measured by the relaxation technique down to $T$ = 0.35 K in a QD Physical Property Measurement System (PPMS) with a $^3$He option. Four-probe ac resistivity measurements were performed in a QD PPMS with the current flowing perpendicular to the $c$-axis. Hall resistivity measurements were performed in a four-wire geometry, where the current was flowing in the $ab$-plane ($I \parallel ab$) and the magnetic field was applied to the $c$-axis ($H \parallel c$). In order to remove magnetoresistance effects due to voltage-contact misalignment the magnetic field directions were reversed. Samples for resistivity measurements were prepared by attaching Pt-wires using silver paste and silver epoxy. The contact resistance for both silver paste and silver epoxy was high, of the order of $\sim$ 20 to $\sim$ 100 $\Omega $ at room temperature. In order to avoid sample heating due to the high contact resistance, a current of less than 10 $\mu$A was applied below 10 K. Note that, due to the sample heating, it was not possible to measure the sample resistance below 1 K.

\section{Results}

\begin{figure}
\centering
\includegraphics[width=1\linewidth]{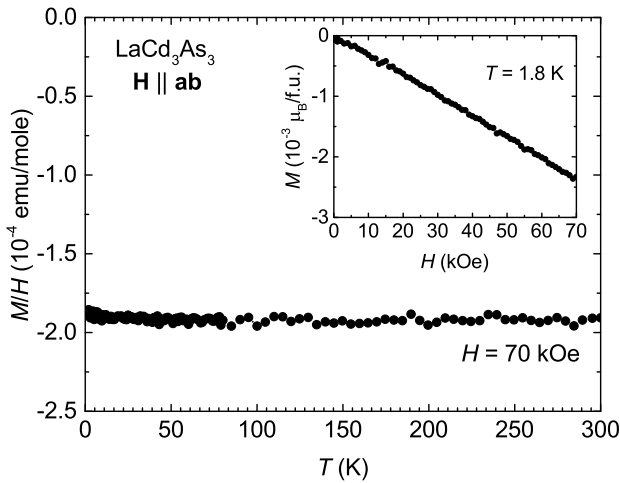}
\caption{Magnetic susceptibility of LaCd$_{3}$As$_{3}$ at $H$ = 70 kOe. Inset shows the magnetization as a function of field at $T$ = 1.8 K.}
\label{Fig2}
\end{figure}

The magnetic properties of LaCd$_{3}$As$_{3}$ are consistent with a diamagnetic compound. Figure \ref{Fig2} shows the temperature dependent magnetic susceptibility, $\chi (T) = M/H$, at $H~=~70$~kOe, applied in the $ab$-plane. The magnetic susceptibility depends weakly on temperature with a magnitude of $\sim$ -2 $\times$ 10$^{-4}$ emu per formula unit. At $T$ = 1.8 K, the magnetization decreases quasi-linearly with increasing magnetic field up to 70 kOe, as shown in the inset of Fig. \ref{Fig2}.

\begin{figure}
\centering
\includegraphics[width=1\linewidth]{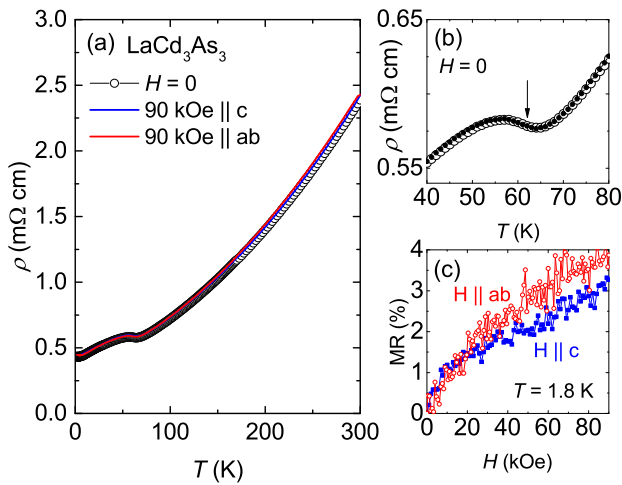}
\caption{(a) Electrical resistivity of LaCd$_{3}$As$_{3}$ at $H$ = 0 and 90 kOe for both $H \parallel ab$ and $H \parallel c$. (b) Zero-field resistivity with increasing (open symbol) and decreasing (closed symbol) temperature between 40 and 80 K. Vertical arrow indicates a peak position in d$\rho (T)$/d$T$. (c) Magnetoresistance at $T~=~1.8$~K for both $H \parallel ab$ and $H \parallel c$.}
\label{Fig3}
\end{figure}

Figure \ref{Fig3} shows the temperature and magnetic field dependent electrical resistivity, $\rho (T,H)$, for LaCd$_{3}$As$_{3}$. The resistivity exhibits metallic behavior with a very slight concave curvature above 70 K. On decreasing temperature $\rho (T)$ exhibits a inflexion point around $T_{0}~\sim~63$~K, determined by the analysis of d$\rho (T)$/d$T$. This feature shows no thermal hysteresis on cooling and warming, as illustrated in Fig. \ref{Fig3}(b). Measurement of the magnetic susceptibility shows no discontinuities or changes in slope around $T_0$. Thus, in the absence of any magnetic signature, it is expected that the anomaly at $T_0$ is related to the opening of a gap in the density of states at the Fermi level, as discussed below. The magnetoresistance (MR), $[\rho (H)-\rho (0)]/\rho (0)$, measured for both $H \parallel ab$ and $H \parallel c$, is small and positive for the entire temperature range as shown in Fig. \ref{Fig3}(a) and (c). The MR shows minimal anisotropy between $H \parallel ab$ and $H \parallel c$.

\begin{figure}
\centering
\includegraphics[width=1\linewidth]{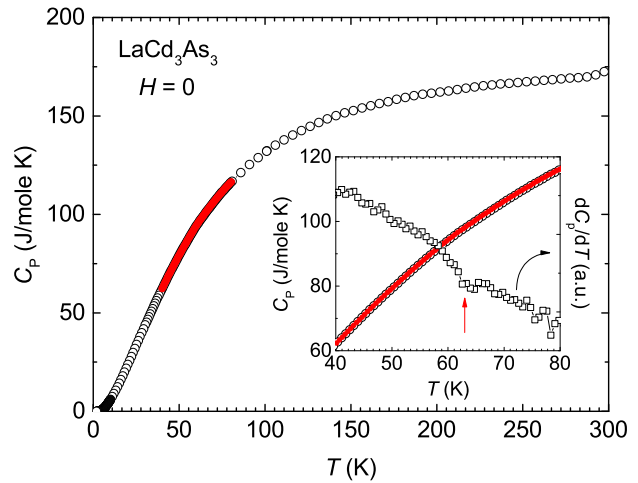}
\caption{Specific heat of LaCd$_{3}$As$_{3}$. The open and closed-symbols represent measurements taken on warming and cooling, respectively. Inset shows an expanded plot around $T_{0}$ (left axis) and the d$C_{p}$/d$T$ (right axis). }
\label{Fig4}
\end{figure}

The temperature dependence of the specific heat for LaCd$_{3}$As$_{3}$ is shown in Fig. \ref{Fig4}. The specific heat reaches $\sim$ 170 J/mole-K at 300 K, which is close to the Dulong-Petit limit and does not show any sharp signature of a phase transition below 300 K.  In addition, no detectable thermal hysteresis within the resolution of our heat capacity measurement has been observed around $T_0$.  However, d$C_{p}$/d$T$ analysis does indicate a subtle slope change at $T_0$ (see the inset of Fig \ref{Fig4}).

\begin{figure}
\centering
\includegraphics[width=1\linewidth]{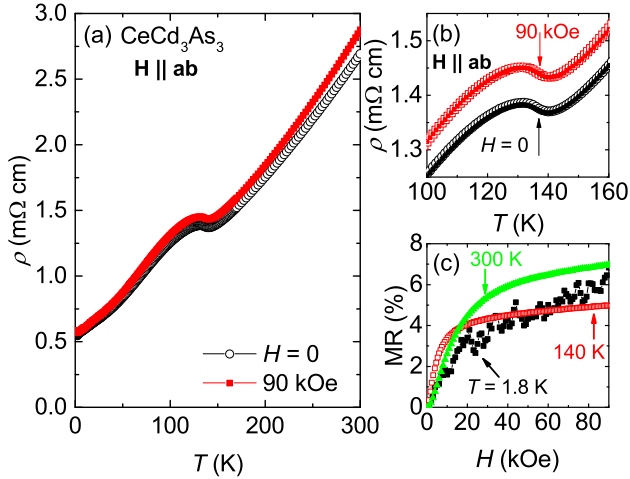}
\caption{(a) Electrical resistivity of CeCd$_{3}$As$_{3}$ at $H$ = 0 and 90 kOe for $H \parallel ab$. (b) Resistivity with increasing (open symbol) and decreasing (closed symbol) temperature between 100 and 160 K. (c) Magnetoresistance at $T$ = 1.8, 140, and 300 K for $H \parallel ab$.}
\label{Fig5}
\end{figure}

Electrical resistivity curves of CeCd$_{3}$As$_{3}$ at $H$ = 0 and 90 kOe for $H \parallel ab$ are displayed in Fig. \ref{Fig5}(a). In zero field, clear metallic behavior is evident at high temperatures. There is an abrupt change in $\rho (T)$ around $T_{0}$ = 136 K, determined by a peak in d$\rho (T)$/d$T$ and indicated by arrows in Fig. \ref{Fig5}(b) for both $H$ = 0 and 90 kOe. An applied magnetic field of 90 kOe does not shift the value of $T_{0}$ and there is no evidence of thermal hysteresis. The resistivity of LaCd$_{3}$As$_{3}$ also shows a similar anomaly around 63~K, which mostly likely has the same origin as the 136~K anomaly seen in CeCd$_{3}$As$_{3}$. We note that the magnetic field dependence of the resistivity cannot be reliably measured for $H \parallel c$ due to the torque. The residual resistivity is $\sim$ 0.52 m$\Omega$ cm, yielding a residual resistivity ratio which is estimated to be $\rho(300 K)/ \rho(2K)$ $\sim$ 5. It should be emphasized that the electrical resistivity does not show behavior typical of the Kondo lattice compounds. Specifically, no maxima or logarithmic upturns resulting from crystalline electric field effects in conjunction with the Kondo scattering of conduction electrons on magnetic Ce ions is observed. Further, in general, typical Kondo lattice compounds show a negative MR at low temperatures \cite{bauer91}. By contrast, a small positive MR is observed in CeCd$_{3}$As$_{3}$ for the entire temperature range measured, as shown in Fig. \ref{Fig5}(a) and (c). Basically, except for the temperature of the anomaly at $T_{0}$, the resistivity as a function of temperature of CeCd$_{3}$As$_{3}$ is the same as that of LaCd$_{3}$As$_{3}$.

\begin{figure}
\centering
\includegraphics[width=1\linewidth]{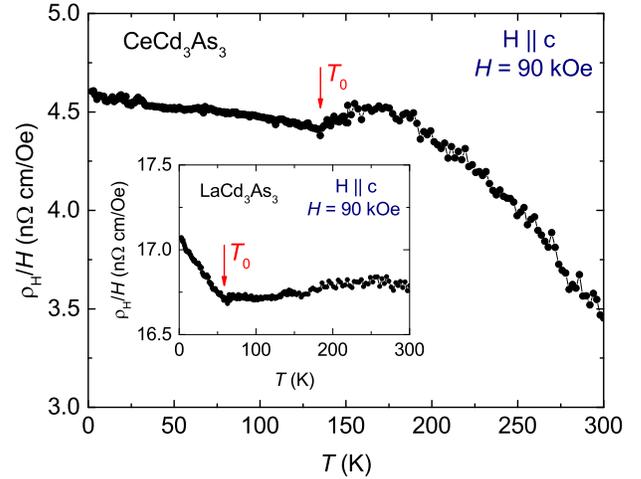}
\caption{Hall coefficient $R_{H} = \rho_{H}/H$ of CeCd$_{3}$As$_{3}$ at $H = 90$~kOe for $H \parallel c$. Inset shows $R_{H}$ of LaCd$_{3}$As$_{3}$ at $H = 90$~kOe for $H \parallel c$.}
\label{Fig11}
\end{figure}

A relatively large value of resistivity for both La- and Ce-compounds suggests a low carrier concentration in these systems. To estimate carrier concentration, the Hall resistivity $\rho_{H}$ has been measured as a function of temperature at fixed magnetic field of 90 kOe along the $c$-axis. To avoid problem with sample rotation, a very thin and small CeCd$_{3}$As$_{3}$ sample was glued to the thermal bath using GE-varnish. The temperature dependence of the Hall coefficient, $R_{H} = \rho_{H}/H$, is plotted in Fig.~\ref{Fig11} for CeCd$_{3}$As$_{3}$ and the inset for LaCd$_{3}$As$_{3}$. The $R_{H}$ curves of both compounds clearly indicate a slope change at the phase transition temperature $T_{0}$. The positive sign of $R_{H}$ indicates that the electrical transport is dominated by hole-like charge carriers. Based on the one-band model approximation, the carrier density at 300~K is estimated to be $\sim 6.7\times10^{-4}$ carriers per formula unit (f.u.) for LaCd$_{3}$As$_{3}$ and $\sim 3.3\times10^{-3}$ carriers per f.u. for CeCd$_{3}$As$_{3}$, indeed confirming the low carrier density of these systems. The estimated carrier density of CeCd$_{3}$As$_{3}$ is comparable to that of CeCd$_{3}$P$_{3}$ ($\sim$0.002 carriers per f.u.) \cite{lee19} and smaller than that of YbAl$_{3}$C$_{3}$ ($\sim$0.01 carriers per f.u.) \cite{ochiai07}.

\begin{figure}
\centering
\includegraphics[width=1\linewidth]{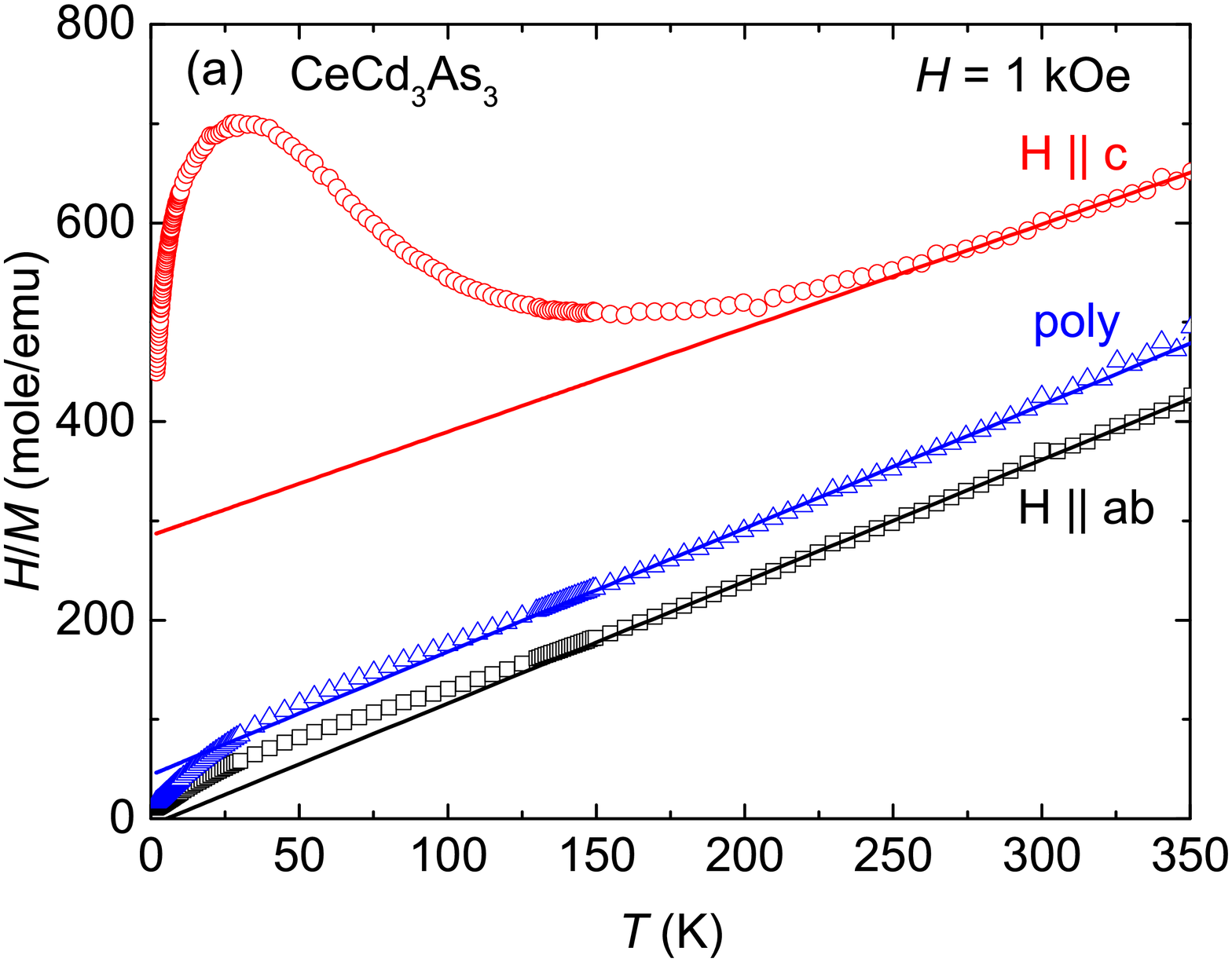}
\includegraphics[width=1\linewidth]{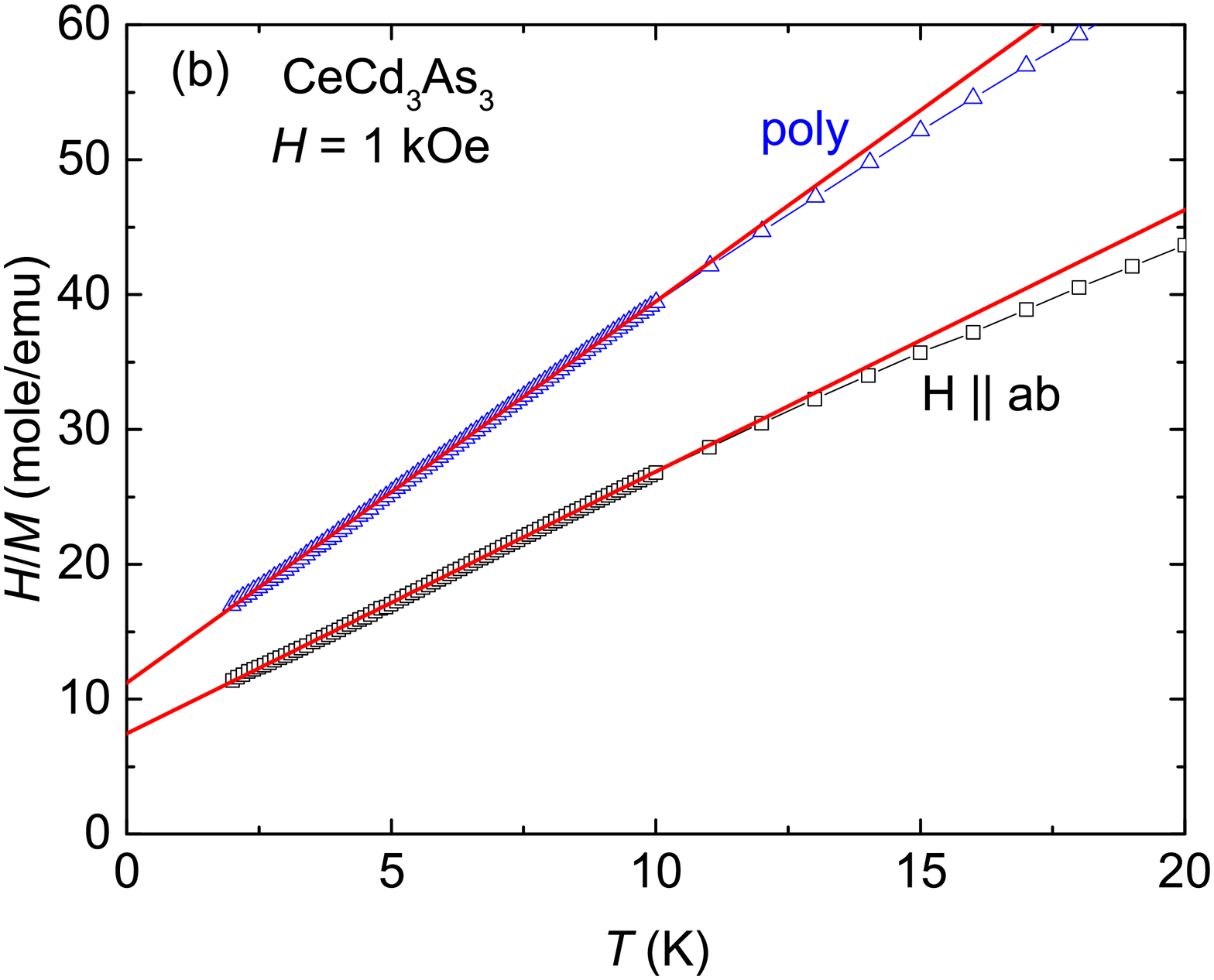}
\caption{(a) The inverse magnetic susceptibility $1/\chi(T) = H/M$ of CeCd$_{3}$As$_{3}$ at $H$ = 1 kOe for $H \parallel ab$, $H \parallel c$, and the polycrystalline average. (b) $1/\chi(T)$ at $H$ = 1 kOe for $H \parallel ab$ and the polycrystalline average below 20~K. Solid lines are from fits of the Curie-Weiss law to the data.}
\label{Fig6}
\end{figure}

The inverse magnetic susceptibility $1/\chi (T) = H/M$ curves of CeCd$_{3}$As$_{3}$ are shown in Fig. \ref{Fig6} (a), for applied magnetic fields both parallel ($\chi _{c}$) and perpendicular ($\chi_{ab}$) to the $c$-axis, which are consistent with an earlier report \cite{liu16}. Here it should be emphasized that no signature for the phase transition at $T_{0}$ is detected in either curve, although it is very clear in $\rho (T)$. The magnetic susceptibility of the polycrystalline average is estimated as $\chi_{poly} = (2/3)\chi_{ab} + (1/3) \chi_{c}$ over the entire temperature range. Note that $\chi _{c}$ has a broad peak around 150~K, which is consistent with an earlier report \cite{liu16}. The inverse magnetic susceptibility $1/\chi (T)$ curves follow a linear temperature dependence above 200~K for $H \parallel ab$ and 275~K for $H \parallel c$. Therefore, to estimate the effective moment $\mu_{eff}$ and Curie-Weiss temperature $\theta_{p}$, we fit the curve by the Curie-Weiss law, $\chi (T) = C/(T-\theta_{p})$, from the linear region of the $1/\chi (T)$ curves: $\mu^{ab}_{eff}$ = 2.55~$\mu_{B}$ and $\theta^{ab}_{p}$ = 5.3~K for $\chi_{ab}$; $\mu^{c}_{eff}$ = 2.76~$\mu_{B}$ and $\theta^{c}_{p}$ = -273~K for $\chi_{c}$; and $\mu^{poly}_{eff}$ = 2.54~$\mu_{B}$ and $\theta^{poly}_{p}$ = -35.4~K for $\chi_{poly}$. Note that it would be necessary to measure $\chi (T)$ above 350 K to determine the $\mu_{eff}$ and $\theta_{p}$ more reliably along the $c$-axis. The obtained effective moment agrees well with the corresponding free Ce$^{3+}$ ion value $(\mu_{eff}$ = 2.54 $\mu_{B})$. For comparison and to avoid potential systematic effects due to the changes in the thermal population of CEF energy levels, $\mu_{eff}$ and $\theta_p$ are also estimated by fitting the magnetic susceptibility curves below 10~K as shown in Fig. \ref{Fig6} (b): $\mu^{ab}_{eff}$ = 2.02~$\mu_{B}$ and $\theta^{ab}_{p}$ = -3.8~K for $\chi_{ab}$ and $\mu^{poly}_{eff}$ = 1.68~$\mu_{B}$ and $\theta^{poly}_{p}$ = -3.95~K for $\chi_{poly}$. Note that the value of the low temperature $\theta^{ab}_{p}$ is close to the value (-4.5 K) of previous report \cite{liu16}. The $1/\chi (T)$ for $H \parallel c$ shows no linear temperature dependence below the maximum around $\sim$30 K. It should be noted that due to the strong CEF contribution in the magnetic susceptibility curves the obtained $\theta_{p}$ values from the Curie-Weiss law can't be used to estimate the size of the magnetic interactions.

\begin{figure}
\centering
\includegraphics[width=1\linewidth]{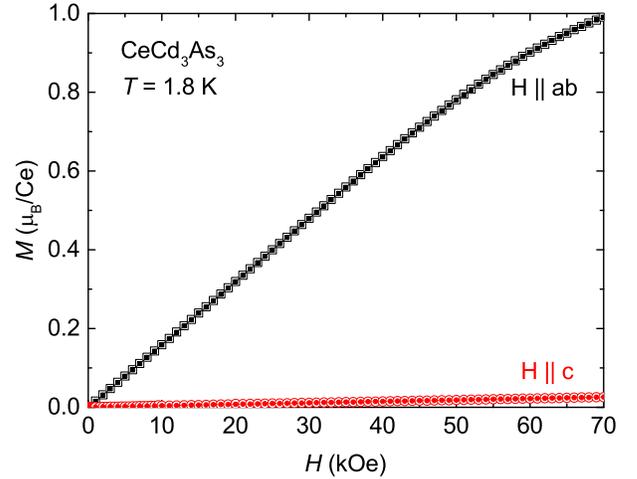}
\caption{Magnetization isotherms for $H \parallel ab$ and $H \parallel c$ at $T$ = 1.8 K. Open and closed-symbols represent up sweep and down sweep, respectively.}
\label{Fig7}
\end{figure}

The field dependent magnetization $M(H)$ up to 70 kOe at 1.8 K is shown in Fig. \ref{Fig7}; it indicates a large anisotropy between $H \parallel ab$ and $H \parallel c$ with an easy axis within the $ab$-plane. For $H \parallel c$, $M(H)$ increases linearly up to 70 kOe. On the other hand, $M(H)$ along $H \parallel ab$ increases linearly up to $\sim$ 40 kOe and tends to roll over slightly at higher magnetic field, reaching the value of $\sim$ 1 $\mu_B$/Ce$^{3+}$ at 70 kOe. The value of $M(H)$ at 70 kOe is smaller than the theoretical value of 2.14 $\mu_B$ for the saturated moment of free Ce$^{3+}$ ions, probably due to the CEF effect. Note that the value of $M(H)$ at 70 kOe is slightly smaller than that of Ref. \cite{liu16}.

\begin{figure}
\centering
\includegraphics[width=1\linewidth]{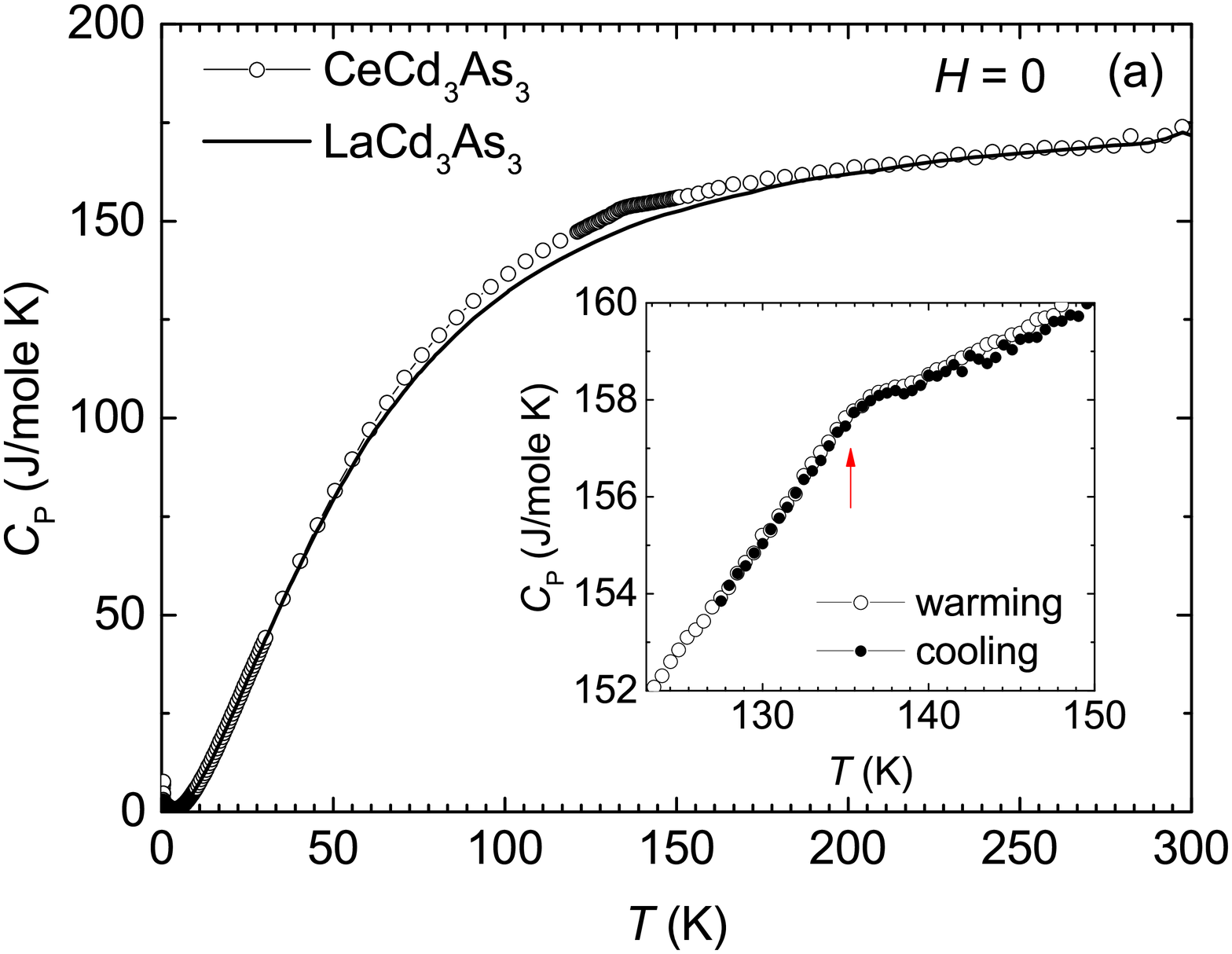}
\includegraphics[width=1\linewidth]{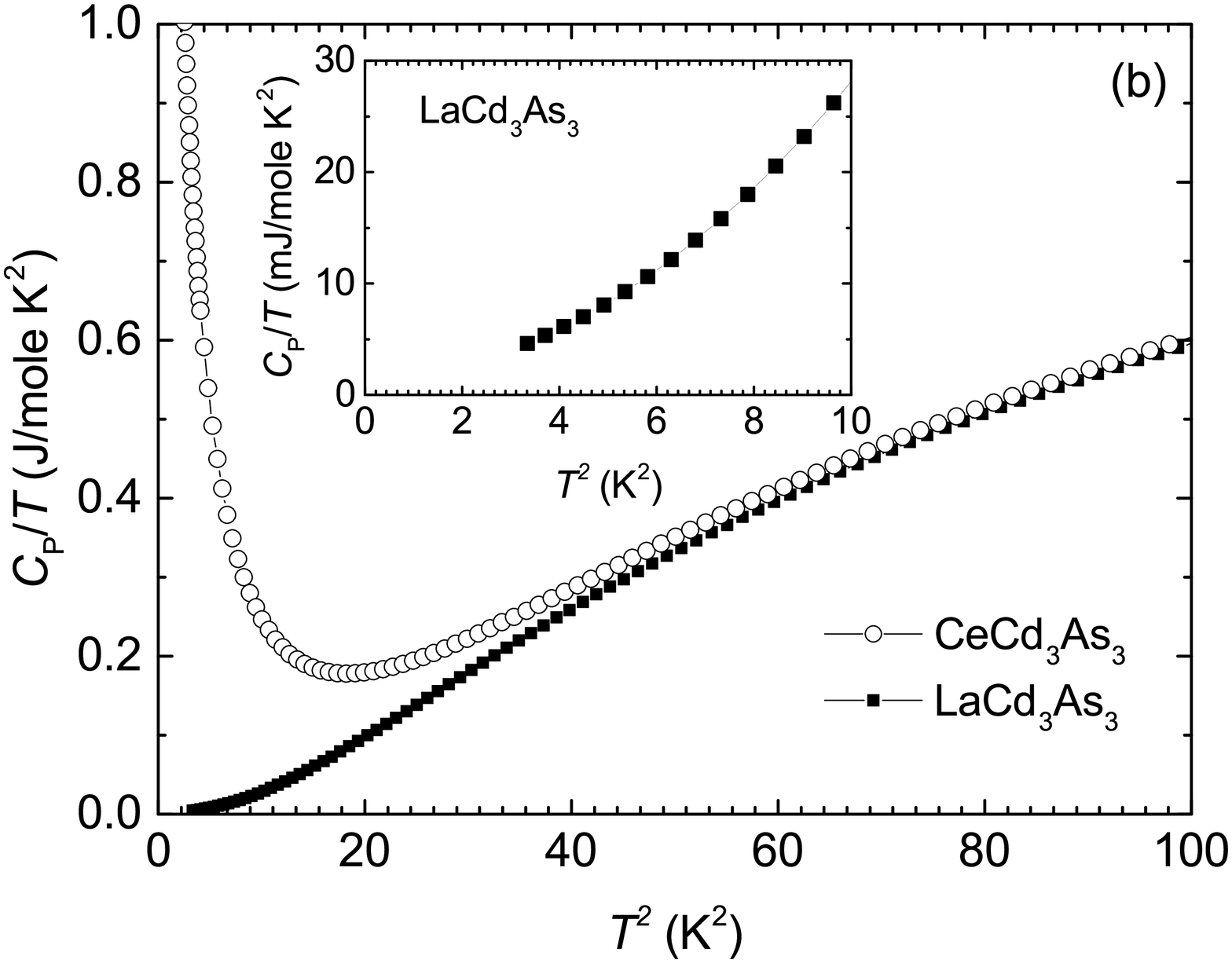}
\caption{(a) $C_{p}$ for both CeCd$_{3}$As$_{3}$ and LaCd$_{3}$As$_{3}$. Inset shows specific heat around $T_{0}$. The open- and closed-circles represent measurements taken on warming and cooling the temperature, respectively. (b) $C/T$ vs $T^{2}$. Inset shows the specific heat of LaCd$_{3}$As$_{3}$ at low temperatures.}
\label{Fig8}
\end{figure}

\begin{figure}
\centering
\includegraphics[width=1\linewidth]{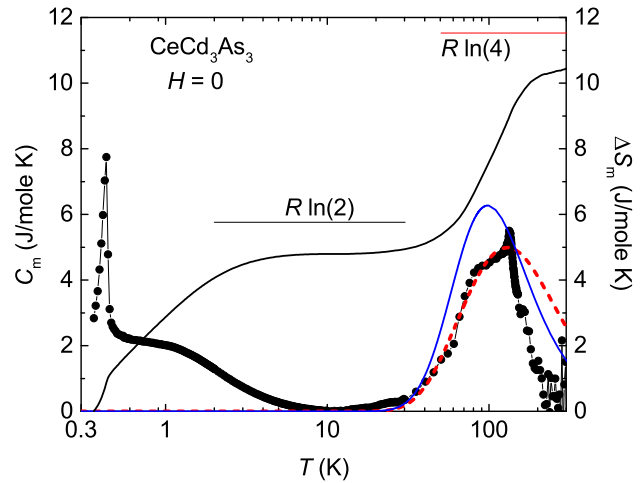}
\caption{Magnetic part of specific heat $C_{m}$ (left axis) and magnetic entropy $S_{m}$ (right axis). The dashed-line represents the Schottky contribution based on a three-doublets Schottky contribution with $\Delta_{1}$ = 240 K and $\Delta_{2}$ = 520 K. The solid-line represents the Schottky contribution base on Ref. \cite{banda18}, where $\Delta_{1}$ = 241~K $\Delta_{2}$ = 282~K}
\label{Fig9}
\end{figure}

Figure \ref{Fig8}(a) illustrates the temperature dependence of the specific heat $C_{p}$ of single crystalline CeCd$_{3}$As$_{3}$ together with that of LaCd$_{3}$As$_{3}$ for comparison. On cooling, $C_{p}$ of CeCd$_{3}$As$_{3}$ shows a slope change at $T_{0}$ $\sim$136~K (see inset) and evidence of a $\lambda$-like anomaly at 0.42 K (see Fig. \ref{Fig9}). No thermal hysteresis is observed for either anomaly. Note that no anomaly indicative of any magnetic ordering or electronic structure change was reported in an earlier specific heat measurement above 0.6 K~\cite{liu16}. Figure \ref{Fig8}(b) shows specific heat curves below 10 K for both CeCd$_{3}$As$_{3}$ and LaCd$_{3}$As$_{3}$, plotted as $C/T$ vs $T^{2}$. For the LaCd$_{3}$As$_{3}$ compound, since the specific heat curve does not follow $C (T) = \gamma T + \beta T^{3}$ at low temperatures (inset), neither $\gamma$ nor the Debye temperature $\Theta _D$ can be reliably obtained. Thus, the value of $\gamma $ is estimated by a linear extrapolation to zero temperature of the $C (T)/T$ versus $T^{2}$ curve below 2.2~K. The estimate of $\gamma$ $\sim$ 0 reflects either a small electronic contribution or a low carrier density. Note that the $C (T)/T$ value at 1.8 K is $\sim$ 3.3 mJ/mol-K$^2$. For CeCd$_{3}$As$_{3}$, due to the magnetic ordering ($\lambda$-like peak) at low temperatures, the $C(T)/T$ versus $T^{2}$ plot doesn't show a clear linear temperature dependence. Therefore, again, $\gamma$ and $\Theta _D$ cannot be reliably extracted by this analysis.

The magnetic contribution to the specific heat of CeCd$_{3}$As$_{3}$, $C_{m}$ = $C_p$(CeCd$_{3}$As$_{3}$) - $C_p$(LaCd$_{3}$As$_{3}$), and the associated magnetic entropy ($S_m$) obtained by integrating $C_{m} /T$ are plotted in Fig. \ref{Fig9}. Because of the magnetic ordering below 0.42 K, the extrapolation of the specific heat to $T$ = 0 cannot be made without ambiguity. Thus, the integration of $C_{m}/T$ has been performed from the lowest accessible temperature of 0.36 K. This will underestimate the total magnetic entropy, especially at low temperatures. Below 10 K, $C_m$ increases as temperature decreases, and shows a tendency towards saturation.  The observed broad background above 0.6 K is consistent with the earlier specific heat measurement \cite{liu16}.  A $\lambda$-like peak in $C_{m}$ indicative of magnetic ordering develops as the temperature is further reduced ($T_{N}$ = 0.42 K). At $T_{N}$, $\sim$ 30$\%$ of $R$ln(2) of $S_{m}$ is released. As seen from Fig. \ref{Fig9}, $S_m$ approaches about $\sim$ 5 J/mol-K around 5 K, which is smaller than $R$ln(2). When the underestimated entropy below 0.36 K ($\sim$ 1 J/mol-K) is taken into account, we believe the ideal value of this entropy is $R$ln(2), consistent with the twofold degree of freedom of the Kramers doublet ground state resulting from CEF splitting of the isolated ion energy levels given trigonal symmetry. Similarly, considering the underestimated entropy below 0.36 K, roughly 40$\%$ of $R$ln(2) entropy is recovered at $T_{N}$.

As shown in Fig. \ref{Fig9}, there is a change in the slope of $C_{m}$ as a function of temperature around 60~K and a peak around 130~K on top of the Schottky-type broad peak around 100~K. These anomalies are attributed to the phase transition observed in both LaCd$_{3}$As$_{3}$ at $T_{0}$ = 63 K and CeCd$_{3}$As$_{3}$ at $T_{0}$ = 136 K. However, the magnitude of the anomalies is very small, indicating only a small entropy change at $T_{0}$. The entropy recovers to a value of $R\ln(4)$ around 200~K, consistent with the population of an excited doublet in the CEF energy level scheme \cite{banda18}. The dashed line in Fig. \ref{Fig9} represents a Schottky contribution to the specific heat. The analysis, assuming three doublets, yields energy level splittings of $\Delta_{1}$ = 240 K and $\Delta_{2}$ = 520 K between the ground state, first and second excited states, respectively. Note that energy level splittings between the ground states and first and second excited doublets of $\Delta_{1}$ = 241 K and $\Delta_{2}$ = 282 K, respectively, obtained by the analysis of magnetic susceptibility measurements have been reported in Ref. \cite{banda18}. Although there is a discrepancy, it is sufficient to establish that the ground-state CEF doublet is well separated from the first-excited doublet.

\begin{figure}
\centering
\includegraphics[width=1\linewidth]{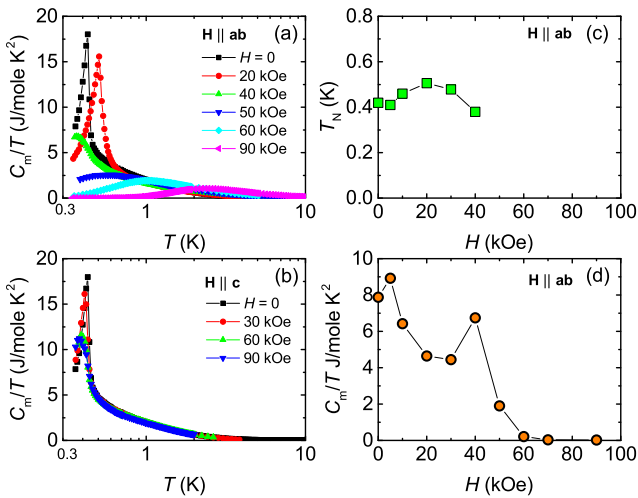}
\caption{CeCd$_{3}$As$_{3}$, (a) $C_{m}/T$ below 10 K at various fields applied in the $ab$-plane ($H \parallel ab$). (b) $C_{m}/T$ below 10 K for $H \parallel c$. (c) $T_{N}$ as a function of field for $H \parallel ab$. (d) $C_{m}/T$ at $T$ = 0.36 K for $H \parallel ab$.}
\label{Fig10}
\end{figure}

The $C_{m}/T$ curves at different magnetic fields for both $H \parallel ab$ and and $H \parallel c$ are plotted in Fig. \ref{Fig10}(a) and (b), respectively. For $H \parallel ab$, the magnetic transition temperature $T_{N}$ in zero field shifts slightly to higher temperature up to 20 kOe. On increasing the field further up to $H$ = 40 kOe, the peak position shifts to lower temperature as shown in Fig. \ref{Fig10}(c). Finally, for $H$ = 50 kOe, the magnetic ordering is suppressed below accessible temperatures. Since the magnetic ordering is suppressed by external magnetic fields, it is expected that the 0.42 K phase transition is not related to ferromagnetic but rather to antiferromagnetic ordering. In the high field region, a broad peak structure in $C_{m}/T$ develops for $H$ $>$ 40 kOe and moves to higher temperature as magnetic field increases. For $H > 40$ kOe, the evolution of the value of $\gamma $ as a function of field can be roughly inferred from $C_{m}/T$ at base temperature. As shown in Fig. \ref{Fig10}(d), for $H$ $>$ 60 kOe, $C_{m}/T$ is quickly suppressed. For $H \parallel c$, the low temperature $C_{m}$ is not sensitive to the magnetic field as shown in Fig. \ref{Fig10}(b) and $T_{N}$ remains almost the same even up to 90 kOe, consistent with strong magnetic anisotropy.

\section{Discussion}

At high temperatures the resistivity of LaCd$_{3}$As$_{3}$ clearly shows an anomaly at $T_{0}$ = 63 K, which mostly likely has the same origin as the 136 K anomaly seen in CeCd$_{3}$As$_{3}$. The feature is reminiscent of that associated with charge density wave (CDW) and spin density wave (SDW) order. Since the phenomenon arises in the resistivity of both compounds and does not appear in the magnetic response, its origin is not ascribed to a magnetic transition. Rather, it is reasonable to assume that this anomaly is related to a change in electronic structure or a structural phase transition. In the latter case, the specific heat must show evidence of a very narrow peak and the resistivity an abrupt jump as a consequence of a first order phase transition. Given the small feature in the specific heat data of CeCd$_{3}$As$_{3}$ and LaCd$_{3}$As$_{3}$, we speculate rather that the high temperature anomaly in the resistivity is due to CDW ordering weakly coupled to the lattice. We expect that the resistivity anomaly at high temperatures is a common characteristic of the $R$Cd$_3$As$_3$ ($R$ = La, Ce, and Pr) family of compounds. If our conjecture is correct, $\rho (T)$ of PrCd$_{3}$As$_{3}$ will show a similar anomaly at higher temperature ($T_{0} >$ 136 K), moving $T_{0}$ to higher temperature with decreasing unit cell volume.  Such an investigation is underway.  Note that a structural phase transition has been reported in a number of members of the $R$Al$_{3}$C$_{3}$ ($R$ = rare-earth) \cite{ochiai10} and $R$Cd$_{3}$P$_{3}$ \cite{lee19} family of compounds. In the case of YbAl$_{3}$C$_{3}$, a structural phase transition at 77 K from a  high temperature hexagonal to a low temperature orthorhombic phase has been confirmed \cite{ochiai07, matsumura08, dhar08}, though the distortion is very minor. To further elucidate the nature of the transition detected in $R$Cd$_{3}$As$_{3}$, high resolution XRD measurements below $T_{0}$ are necessary. Since the CDW order is predominantly driven by Fermi surface nesting, it is especially sensitive to pressure induced changes in the electronic structure. Therefore, performing electrical resistivity measurements under hydrostatic pressure will also clarify the nature of the transition at $T_{0}$.

We now turn to a discussion of the low temperature specific heat of CeCd$_{3}$As$_{3}$. The magnetic part of the specific heat $C_{m}$ increases below 10~K and shows a tendency to saturate below 1~K, while magnetic ordering occurs at 0.42~K. Although the specific heat clearly indicates magnetic ordering, which can be suppressed for $H$ $>$ 40 kOe along $H \parallel ab$, the order parameter is currently unknown. The associated ordered spin structure below $T_{N}$ could be investigated using neutron diffraction with an isotopically enriched sample to overcome the large neutron absorption cross-section of $^{113}$Cd. Although the Sommerfeld coefficient $\gamma$ of CeCd$_{3}$As$_{3}$ cannot be reliably estimated due to the magnetic ordering and the Schottky contributions, there is a large electronic specific heat contribution at low temperatures. As is usual for other Ce-based systems, the increase in $C_m$ below 10 K may be interpreted in terms of a large value of $\gamma$ associated with the Kondo effect. However, the resistivity curve indicates no Kondo lattice behavior, as would be evidenced by $\log(T)$-dependence or coherence behavior. It would be necessary to measure resistivity below 1 K to see any Kondo contribution. A measurement of the specific heat of CeCd$_{3}$As$_{3}$ down to 20 mK will also clarify whether heavy quasiparticles survive at low temperatures. If the value of $C_{m}/T$ is small much below the $T_{N}$, the large $C_{m}$ below 10~K can be attributed to other effect such as magnetic fluctuations.

We speculate that the absence of a response in the resistivity typical of a Kondo lattice is due to the low carrier density. Usually in the Kondo lattice system, a substantial concentration of conduction carriers are thought to be necessary for the formation of a Kondo-singlet. The low carrier density can be inferred from the unusually high resistivity values of CeCd$_{3}$As$_{3}$ and LaCd$_{3}$As$_{3}$, of the order of 10-100 times larger at room temperature than typical Ce-based intermetallic systems. In addition, the Hall coefficient measurements for La- and Ce-based samples confirm the extremely low carrier density of these systems. We note that the low carrier density is a common feature of this family of materials, where the resistivity measurements for polycrystalline CeZn$_{3}$P$_{3}$ and CeCd$_{3}$P$_{3}$ show semiconducting behavior \cite{yamada10, higuch16} and a low carrier concentration in single crystal $R$Al$_{3}$C$_{3}$ \cite{ochiai07} and CeCd$_{3}$P$_{3}$  \cite{lee19} has been suggested from resistivity and Hall coefficient measurements. Because it depends on carriers near as well as deep inside the Fermi sea \cite{CastroNeto2000}, it is expected that the strength of the RKKY interaction will be weakened by reducing the carrier density.  Since both the RKKY and Kondo effect are mediated by the conduction electrons, other short-range interactions (such as super-exchange) may be made progressively more important by reducing the carrier density. It is therefore not unreasonable to consider both short-range and RKKY interactions to explain the magnetic properties of CeCd$_{3}$As$_{3}$.

The question arises whether geometric magnetic frustration plays any role in CeCd$_{3}$As$_{3}$. For a well localized moment (non-hybridized) system in the absence of frustration, except for the low dimensional systems, $\theta_{p}$ should be comparable to the magnetic ordering temperature. Analyzing the frustration factor of CeCd$_{3}$As$_{3}$, $f = \theta_{p}/T_{N}$, one may suspect it is indeed relevant, since $T_{N}$ = 0.42 K is clearly much smaller than the $\theta_{p}$ values extracted from both high and low temperature Curie-Weiss fit. It has been shown theoretically that  a broad maximum in the specific heat can develop in square and anisotropic triangular lattice systems, depending on the ratio of in-plane exchange interactions \cite{schmidt17}. Hence it is possible to qualitatively explain the large value of $C_{m}/T$ below 10 K observed in CeCd$_{3}$As$_{3}$ without invoking Kondo physics. Such a broad maximum in the specific heat has also been observed in many insulating triangular lattice compounds such as YbMgGaO$_{4}$ \cite{xu16}. It should be emphasized that a large linear specific heat coefficient ($\gamma$ $\sim$ 200 mJ/mol-K$^{2}$) caused by charge ordering rather than due to the Kondo effect is observed in the low carrier density system Yb$_{4}$As$_{3}$ \cite{kohgi97}. The evolution of magnetic entropy is also consistent with a strong effect due to frustration in CeCd$_3$As$_3$ compound. If there is no frustration, the $R\ln(2)$ ground state entropy must be recovered at $T_{N}$. On CeCd$_3$As$_3$ only 40$\%$ of $R\ln(2)$ entropy is recovered by the ordering temperature (Fig. \ref{Fig9}) and the remaining contribution by $\sim 10$~K. This indicates that a sizable contribution associated with short-range correlations exists above $T_{N}$, as often observed in insulating magnetically frustrated compounds and low dimensional systems \cite{collins97}.

A common feature of 1-D and 2-D systems, either calculated theoretically or observed experimentally, is the development of a broad maximum in the magnetic susceptibility at the temperature comparable with the energy scale of the exchange interaction $J_{ex}$ \cite{zheng05, Shimizu2003}. For the CeCd$_3$As$_3$ compound, the lattice planes containing Ce$^{3+}$ ions are separated $\sim$~10.7~\AA~by Cd and As atoms, far larger than the in-plane separation. It has been suggested \cite{liu16} that CeCd$_3$As$_3$ is a very strong Ising-type system, where the maximum in $\chi_{c}$ is due to a large anisotropy of the interaction between $ab$-plane and $c$-axis. However, the broad maximum in $\chi_{c}$ centered around 150 K (Fig. \ref{Fig6}) cannot be indicative of the $c$-axis exchange interaction - the temperature scale is unreasonably high. In addition, the magnetic susceptibility curves have been analyzed by using a point charge CEF model with trigonal point symmetry and showed a good agreement \cite{banda18}. Thus, the large negative value of the $\theta^{c}_{p}$ $\simeq$ $-$273 K and the broad peak observed in $\chi_{c}$ are likely dominated by CEF effects. The magnetic part of the specific heat shows evidence of a broad Schottky peak centered around 100~K, which is consistent with the magnetic susceptibility data. Therefore, caution should be exercised in interpreting the values of $\theta_{p}$ extracted from the high temperature magnetic susceptibility measurement. Alternatively, extracting $\theta_{p}$ from data at lower temperatures runs the risk of inadvertently including correlation effects. In principle, a negative value of the low temperature $\theta^{ab}_{p} \simeq -3.8$~K for $\chi_{ab}$ would indicate antiferromagnetic in-plane coupling in CeCd$_{3}$As$_{3}$. However, this would also include several in-plane interactions in the anisotropic triangular lattice \cite{schmidt15}. Since an anomaly in the resistivity and specific heat occurs at $T_{0}$, it may be accompanied by slight displacement of the atoms. This displacement will cause the deformation of the triangular lattice, altering distances between Ce-ion and its nearest neighbor Ce-ions. Although a large volume change is not expected at $T_{0}$, nonetheless the strengths of the exchange coupling between nearest neighbor Ce-ions would be altered, resulting in changes in the interactions on the anisotropic triangular lattice.

Finally we discuss the evolution of the magnetic ordering with external magnetic field \cite{ye17}.  As shown in Fig. \ref{Fig10}, the specific heat measurement along $H \parallel ab$ clearly indicates that the magnetic ordering can be tuned toward $T$~$\rightarrow$~0 and possibly a quantum critical point \cite{gegenwart08}. Theoretically, quantum fluctuations may be particularly prevalent near the phase boundaries, such that spin liquid ground states are stabilized within certain ranges of the in-plane anisotropy parameter \cite{li16}. Developments in describing quantum phase transitions in the presence of frustration have been recently reviewed \cite{vojta17}. It would be of interest to investigate such a scenario by measuring magnetization and specific heat of CeCd$_{3}$As$_{3}$ below 0.36~K. In addition, the origin of the sizable specific heat contribution at low temperature in CeCd$_{3}$As$_{3}$ may be explored by tuning the system to either a more metallic or insulating regime. This changes the fine balance of the characteristic energy scales associated with the Kondo, RKKY, and short-range interactions. One possible avenue is through pressure. Since CeZn$_{3}$P$_{3}$ exhibits semiconducting behavior, it would be illuminating to undertake a doping study (chemical pressure), replacing the Cd/As by Zn/P in CeCd$_{3}$As$_{3}$ to make the system more insulating. By contrast, resistivity measurements of CeZn$_{3}$P$_{3}$ reveal that the semiconducting behavior is strongly suppressed by applying pressure \cite{yamada10}. Thus, an experiment under hydrostatic pressure may provide a way of tuning the system towards a more metallic state. In this scenario, it would be interesting to look for the recovery of Kondo lattice behavior.

\section{Summary}

In summary, the intermetallic systems $R$Cd$_{3}$As$_{3}$ ($R$ = La and Ce), which crystallize in the hexagonal ScAl$_{3}$C$_{3}$-type structure with one apparent rare-earth triangular sublattice, have been investigated by the measurement of thermodynamic and transport properties. The common features of these compounds are the metallic behavior with an anomaly observed at high temperatures, probably associated with electronic structure changes at $T_{0}$. The absence of a significant Kondo effect in CeCd$_3$As$_3$ is demonstrated by the temperature dependence of the resistivity. The magnetic susceptibility measurements show that the magnetic moments of Ce-ions are well localized in the CeCd$_3$As$_3$ compound. The magnetization exhibits a strong magnetic anisotropy as expected from the layered crystal structure. In zero field the specific heat measurement for CeCd$_3$As$_3$ shows a sharp $\lambda$-like anomaly at $T_{N}$ = 0.42 K as a signature of magnetic ordering. On application of a magnetic field along $H \parallel c$, $T_{N}$ remains largely unchanged up to 90 kOe. By contrast, applying the magnetic field along $H \parallel ab$ induces a subtle increase in $T_{N}$ up to 20 kOe, followed by a gentle decrease up to 40 kOe. Above 40 kOe, the transition can be suppressed below the lowest temperature accessible. The CeCd$_3$As$_3$ compound is likely characterized by very large magnetic fluctuations due to the short range interaction and the long range magnetic order is suppressed to 0.42 K by the weak RKKY exchange interaction and most probably by magnetic frustration.

\section{Acknowledgements}

We would like to acknowledge H. Park for sample growths. This work was supported by the Canada Research Chairs, Natural Sciences and Engineering Research Council of Canada, and Canada Foundation for Innovation program.

\end{document}